\documentclass[a4paper,12pt]{article}
\usepackage{amsmath}
\usepackage{amssymb}
\usepackage{latexsym}
\usepackage{enumerate}
\usepackage{graphicx}
\usepackage{placeins}
\newcommand{\be}{\begin{eqnarray}}
\newcommand{\ee}{\end{eqnarray}}

\newcommand{\bsub}{\begin{subequations}}
\newcommand{\esub}{\end{subequations}}

\newcommand{\disfrac}[1][2]{\displaystyle\frac}
\begin{document}

\title{Energy distribution of a regular black hole solution in
Einstein-nonlinear electrodynamics}
\author{I. Radinschi$^{\text{*1}}$, F. Rahaman$^{\text{**2}}$, Th. Grammenos$%
^{\text{***3}}$, \and A. Spanou$^{\text{****4}}$ and Sayeedul Islam$^{\text{%
*****5}}$ \\
$^{\text{1}}$Department of Physics \\
``Gh. Asachi'' Technical University, \\
Iasi, 700050, Romania\\
$^{\text{2}}$Department of Mathematics, Jadavpur University,\\
Kolkata 700 032, West Bengal, India\\
$^{\text{3}}$Department of Civil Engineering, \\
University of Thessaly, 383 34 Volos, Greece\\
$^{\text{4}}$School of Applied Mathematics and Physical Sciences,\\
National Technical University of Athens, 157 80, Athens, Greece\\
$^{\text{5}}$Department of Mathematics, Jadavpur University,\\
Kolkata 700 032, West Bengal, India\\
$^{\text{*}}$radinschi@yahoo.com, $^{\text{**}}$rahaman@iucaa.ernet.in,\\
$^{\text{***}}$thgramme@civ.uth.gr, $^{\text{****}}$aspanou@central.ntua.gr,\\
$^{\text{*****}}$sayeedul.jumath@gmail.com}
\date{}
\maketitle

\begin{abstract}
In this work a study about the energy-momentum of a new four-dimensional spherically
symmetric, static and charged, regular black hole solution developed in the
context of general relativity coupled to nonlinear electrodynamics is presented. Asymptotically, this new
black hole solution  behaves as the Reissner-Nordstr\"{o}m
solution only for the particular value $\mu =4$, where $\mu$ is a
positive integer parameter appearing in the mass function of the solution. The calculations are performed by use of the  Einstein,
Landau-Lifshitz, Weinberg and M\o ller energy-momentum complexes. In all the aforesaid prescriptions, the expressions for the energy of the gravitating system considered depend on the mass $M$
of the black hole, its charge $q$, a  positive integer $\alpha$ and the
radial coordinate $r$. In all these pseudotensorial prescriptions the momenta are found to vanish, while the  Landau-Lifshitz and Weinberg prescriptions give the same result for the energy
distribution.  In addition, the limiting behavior of the energy for
the cases $r\rightarrow \infty $, $r=0$ and $q=0$ is studied. The special case $\mu
=4$ and $\alpha =3$ is also examined. We conclude that the Einstein and M\o %
ller energy-momentum complexes can be considered as the most reliable tools for the
study of the energy-momentum localization of a gravitating system.

\noindent \textbf{Keywords}: Energy-Momentum Complexes; Regular Black Holes.\newline
\textit{PACS Numbers}: 04.20.-q, 04.20.Cv, 04.70.Bw
\end{abstract}

\section{Introduction}

Energy-momentum localization plays an important role among the open issues  appeared over the years in general relativity. The difficulty which arises consists in constructing a properly defined  energy density of
gravitating systems. As a result, up today there is no generally accepted satisfactory description
for gravitational energy.

A number of researchers have used different  methods
for the energy-momentum localization. Standard research methods include the use
of different tools such as superenergy tensors \cite{Bel}, quasi-local expressions
\cite{Brown} and the well-known energy-momentum complexes of Einstein \cite{Einstein},
Landau-Lifshitz \cite{Landau}, Papapetrou \cite{Papapetrou}, Bergmann-Thomson \cite{Bergmann}, M\o ller \cite{Moller}, Weinberg \cite{Weinberg}, and Qadir-Sharif \cite{Qadir}. The coordinate system dependence of these computational tools remains the main problem encountered. An alternative to avoid the dependence on coordinates is the tele-parallel theory of gravitation \cite{Moller_1} which has been used in many studies on the calculation of the energy and momentum distribution.

Concerning the pseudotensorial prescriptions, only the M\o ller energy-momentum complex is coordinate independent. Schwarzschild Cartesian coordinates and Kerr-Schild Cartesian coordinates are used to
compute the energy-momentum in the case of the other pseudotensorial
prescriptions. In the last decades, despite the critique on energy-momentum
complexes, their application has yielded physically reasonable results in
the case of many spacetime geometries, in particular for geometries in $(3+1)$, $(2+1)$ and $(1+1)$ dimensions \cite{Virbhadra}. It is worth noticing that there is an agreement between on one hand the Einstein, Landau-Lifshitz, Papapetrou, Bergmann-Thomson, Weinberg and M\o ller prescriptions and, on the other hand,
the quasi-local mass definition introduced by Penrose \cite{Penrose} and developed by
Tod \cite{Tod} for some gravitational backgrounds. It should be stressed that
several pseudotensorial definitions yield the same
results for any metric of the Kerr-Schild class as well as  for
solutions more general than the Kerr-Schild class (see, e.g., \cite{Aguirregabiria}).
Further, one can underline the similarity of some of the
aforementioned results with those obtained by using the tele-parallel theory
of gravitation \cite{Gamal}. Finally, one can also point out the ongoing elaboration of the definition and application of energy-momentum complexes, as well as the attempt for their rehabilitation \cite{Chang}.

The present work has the following structure: in Section 2 we describe the new four-dimensional charged regular black hole solution \cite{Balart} under study. Section 3  contains the presentation of the Einstein, Landau-Lifshitz, Weinberg and M\o ller energy-momentum complexes utilized for the calculations. Section 4 contains  the computations of the energy and momentum distributions. Finally, in the Discussion given in Section 5, we present our results and examine some limiting and particular cases. 
Throughout the paper we use geometrized units $(c=G=1)$, while the signature
chosen is $(+,-,-,-)$. The calculations for the Einstein, Landau-Lifshitz and Weinberg energy-momentum complexes are performed in Schwarzschild Cartesian coordinates. Greek indices range from $0$ to $3$,
while Latin indices run from $1$ to $3$.

\section{A New Regular Black Hole Solution in Einstein-nonlinear
Electrodynamics}

Attempts to avoid the central singularity in the presence of a horizon for a classical black hole solution have started already in the 1960's with Bardeen's regular black hole solution \cite{Bardeen} followed up by a number of works in the subject (see, \cite{Ansoldi} for a review of existing models and many relevant references). The interest in this research area was revived after E. Ay\'on-Beato and A. Garc\'{\i}a \cite{Ayon} studied a regular black hole solution coupled to nonlinear electrodynamics, whereby the Einstein-Hilbert action of general relativity is augmented by the addition to the Lagrangian of a term depending nonlinearly on the electromagnetic field tensor. Since then, a number of very interesting works has been published on this subject \cite{Beato}. In fact, it is worth noticing that the aforesaid authors were the first to establish a connection of Bardeen's regular black hole solution with a nonlinear electrodynamics leading to the current interpretation o!
 f an exact solution supported by a nonlinear magnetic monopole \cite{Garcia}. Recently, L. Balart and E.C. Vagenas \cite{Balart} constructed a family of new spherical, static and charged black hole solutions in the context of Einstein-nonlinear electrodynamics by imposing the following conditions: (i) the weak energy condition has to be
satisfied, (ii)  the energy-momentum tensor has to respect the symmetry $ T_{0}^{0}=T_{1}^{1}$ and, (iii)
the metric must have an asymptotical
behaviour as the Reissner-Nordstr\"{o}m black hole solution.

Starting with a general static and spherically symmetric line element:
\begin{equation}\label{line_element}
ds^2=B(r) dt^2- B^{-1}(r) dr^2-r^2(d\theta^2+\sin^2\theta d\phi^2),
\end{equation}
with $B(r)=1-\frac{2m(r)}{r}$ and by relaxing the third condition, the authors have also constructed a more
general family of charged regular black hole solutions  which do not behave asymptotically as the
Reissner-Nordstr\"{o}m black hole metric. In this case, the mass function
depends on two  integer constant parameters $\mu \geqslant 4$ and $\alpha\geqslant 1$:
\begin{equation}\label{mass_function}
\begin{split}
m(r) &=\frac{r^{3}q^{2}}{6}\frac{\Gamma(\frac{1}{\alpha})\Gamma 
(\frac{\mu }{\alpha })}{\Gamma (\frac{4}{\alpha })\Gamma 
(\frac{\mu -3}{\alpha })}
\left(\frac{6\Gamma (\frac{4}{\alpha })}{\Gamma  (\frac{1}{\alpha})
\Gamma (\frac{\alpha +3}{\alpha })}\frac{M\,}{q^{2}}\right)^{4}\times\\
&\\
&\,_{2}F_{1}\left(\frac{3}{\alpha },\frac{\mu }{\alpha },\frac{3+\alpha }{\alpha }; -(\frac{6\,\Gamma (\frac{4}{\alpha })}{\Gamma (\frac{1}{\alpha })\Gamma (\frac{\alpha +3}{\alpha })}\frac{M\,}{q^{2}}r)^{\alpha}\right),
\end{split}
\end{equation}
where $q$ is the electric charge, $\Gamma$ is the Gamma function, $M$ is the black hole mass and $\,_{2}F_{1}$ is the Gauss hypergeometric function.
Particular cases such as $\mu =5$, $\alpha =2$ and $\mu =6$, $\alpha =3$ lead to the metrics obtained by Bardeen \cite{Bardeen} and Hayward \cite{Hayward}.

The line element (\ref{line_element}) with the mass function (\ref{mass_function}) does not behave asymptotically  as the Reissner-Nordstr\"{o}m black hole solution unless $\mu =4$.  In fact, for $\mu =4$ and $\alpha =3$ in Eq.(\ref{mass_function}) Balart and Vagenas obtained the
asymptotically Reissner-Nordstr\"{o}m solution
\begin{equation}\label{RN_solution}
B(r)=1-\frac{2M}{r}\left[1-\frac{1}{[1+(\frac{2Mr}{q^2 })^3]^{1/3}}\right].
\end{equation}

In the present paper, we will study the, non-asymprotically Reissner-Nordstr\"{o}m, gravitational background obtained for the particular
case $\mu =3$ and $\alpha$ arbitrary. Consequently the metric for the new charged regular black hole solution to be considered in what follows has the line element (\ref{line_element}) with 
\begin{equation}\label{metric_coef_B}
B(r)= 1-\frac{2M}{r}\left(1-\frac{1}{[1+(\frac{2Mr}{q^{2}})^{3}]^{(\alpha -3)/3}}\right).
\end{equation}

\begin{figure}[b!]
\includegraphics[width=84mm]{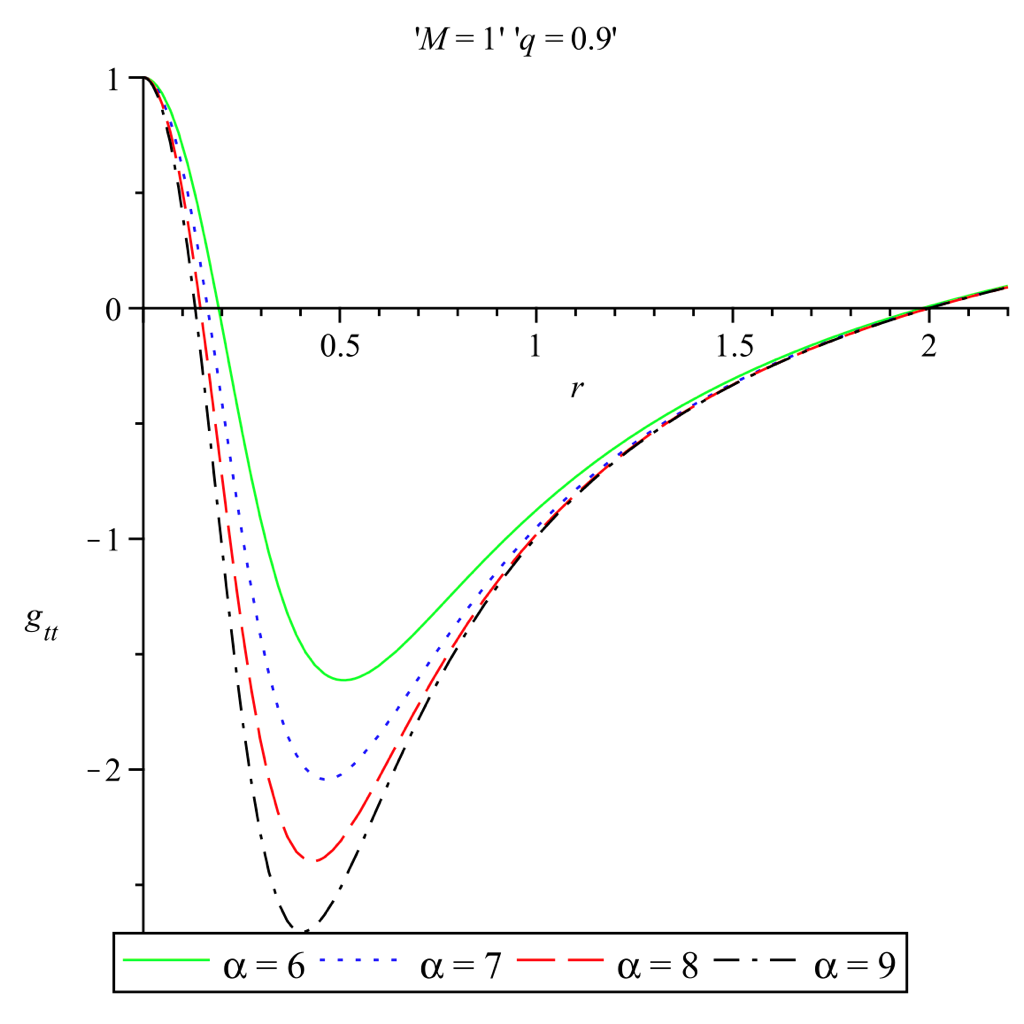}
\caption{Two horizons exist for the charged regular black hole at the points where $g_{tt}$ meets the $r$-axis.}
\label{fig7}
\end{figure}

In fact, this regular black hole solution exhibits two event horizons. By choosing, for example, the numerical values $M=1$ and $q=0.9$, we get  Figure 1 showing the two horizons. One can see that  the position of the inner horizon is shifted towards bigger values of $r$ as the parameter $\alpha$ decreases, while the position of the outer horizon remains unaffected.

\section{Einstein, Landau-Lifshitz, Weinberg and M\o ller Energy-Momentum
Complexes}

The Einstein energy-momentum complex \cite{Einstein} for a $(3+1)$ dimensional
gravitational background is given by the expression 
\begin{equation}
\theta _{\nu }^{\mu }=\frac{1}{16\pi }h_{\nu ,\,\lambda }^{\mu \lambda },
\end{equation}
where the superpotentials $h_{\nu }^{\mu \lambda }$ are given as 
\begin{equation}
h_{\nu }^{\mu \lambda }=\frac{1}{\sqrt{-g}}g_{\nu \sigma }[-g(g^{\mu \sigma
}g^{\lambda \kappa }-g^{\lambda \sigma }g^{\mu \kappa })]_{,\kappa }
\end{equation}
and satisfy the corresponding antisymmetric property 
\begin{equation}
h_{\nu }^{\mu \lambda }=-h_{\nu }^{\lambda \mu }.
\end{equation}
The components $\theta _{0}^{0}$ and $\theta _{i}^{0}$ represent the energy and
the momentum density, respectively. In this prescription the
local conservation law is respected: 
\begin{equation}
\theta _{\nu,\mu }^{\mu }=0.
\end{equation}
Thus, the energy and momentum can be calculated by
\begin{equation}
P_{\mu }=\int \int \int \theta _{\mu }^{0}\,dx^{1}dx^{2}dx^{3}.
\end{equation}
Applying Gauss' theorem, the energy-momentum becomes
\begin{equation}\label{Einstein_P}
P_{\mu }=\frac{1}{16\pi }\int\int h_{\mu }^{0i}n_{i}dS,
\end{equation}
with $n_{i}$ the outward unit normal vector over the surface $dS$.

In the Landau-Lifshitz prescription, the corresponding energy-momentum
complex \cite{Landau} is defined as 
\begin{equation}
L^{\mu \nu }=\frac{1}{16\pi }S_{,\,\rho \sigma }^{\mu \nu \rho \sigma },
\end{equation}
with the Landau-Lifshitz superpotentials: 
\begin{equation}
S^{\mu \nu \rho \sigma }=-g(g^{\mu \nu }g^{\rho \sigma }-g^{\mu \rho }g^{\nu
\sigma }).
\end{equation}
The $L^{00}$ and $L^{0i}$ components are the energy and the momentum density,
respectively. The local conservation law holds:
\begin{equation}
L_{,\,\nu }^{\mu \nu }=0.
\end{equation}
By integrating $L^{\mu \nu }$ over the 3-space one gets for the
energy-momentum: 
\begin{equation}
P^{\mu }=\int \int \int L^{\mu 0}\,dx^{1}dx^{2}dx^{3}.
\end{equation}
By using Gauss' theorem we get 
\begin{equation}\label{LL_P}
P^{\mu }=\frac{1}{16\pi }\int \int S_{,\nu }^{\mu 0i\nu }n_{i}dS=\frac{1}{16\pi }\int \int U^{\mu 0i}n_{i}dS.
\end{equation}

The Weinberg energy-momentum complex \cite{Weinberg} is given as 
\begin{equation}
W^{\mu \nu }=\frac{1}{16\pi }D_{,\,\lambda }^{\lambda \mu \nu },
\end{equation}
where $D^{\lambda \mu \nu }$ represents the corresponding superpotentials: 
\begin{equation}
D^{\lambda \mu \nu }=\frac{\partial h_{\kappa }^{\kappa }}{\partial
x_{\lambda }}\eta ^{\mu \nu }-\frac{\partial h_{\kappa }^{\kappa }}{\partial
x_{\mu }}\eta ^{\lambda \nu }-\frac{\partial h^{\kappa \lambda }}{\partial
x^{\kappa }}\eta ^{\mu \nu }+\frac{\partial h^{\kappa \mu }}{\partial
x^{\kappa }}\eta ^{\lambda \nu }+\frac{\partial h^{\lambda \nu }}{\partial
x_{\mu }}-\frac{\partial h^{\mu \nu }}{\partial x_{\lambda }},
\end{equation}
with 
\begin{equation}
h_{\mu \nu }=g_{\mu \nu }-\eta _{\mu \nu }.
\end{equation}
Here the $W^{00}$ and $W^{0i}$ components represent the energy and
the momentum density, respectively. In the Weinberg prescription the local
conservation law is also respected:
\begin{equation}
W_{,\,\nu }^{\mu \nu }=0.
\end{equation}
The integration of $W^{\mu \nu }$ over the 3-space gives the energy-momentum: 
\begin{equation}
P^{\mu }=\int \int \int W^{\mu 0}\,dx^{1}dx^{2}dx^{3}.
\end{equation}
Applying Gauss' theorem and integrating over the surface of a
sphere of radius $r$, we get the following expression for the energy-momentum distribution:
\begin{equation}\label{W_P}
P^{\mu }=\frac{1}{16\pi }\int \int D^{i0\mu }n_{i}dS.
\end{equation}

The M{\o }ller energy-momentum complex \cite{Moller} is
\begin{equation}
\mathcal{J}_{\nu }^{\mu }=\frac{1}{8\pi }M_{\nu \,\,,\,\lambda }^{\mu
\lambda },
\end{equation}
with the M{\o }ller superpotentials $M_{\nu }^{\mu \lambda }$ 
\begin{equation}
M_{\nu }^{\mu \lambda }=\sqrt{-g}\left( \frac{\partial g_{\nu \sigma }}{\partial x^{\kappa }}-\frac{\partial g_{\nu \kappa }}{\partial x^{\sigma }} \right) g^{\mu \kappa }g^{\lambda \sigma },
\end{equation}
being  antisymmetric:
\begin{equation}
M_{\nu }^{\mu \lambda }=-M_{\nu }^{\lambda \mu }.
\end{equation}
M{\o }ller's energy-momentum complex also satisfies the local conservation
law
\begin{equation}
\frac{\partial \mathcal{J}_{\nu }^{\mu }}{\partial x^{\mu }}=0,
\end{equation}
while $\mathcal{J}_{0}^{0}$  and $\mathcal{J}_{i}^{0} $ are the energy and the momentum density respectively. Thus, the energy-momentum is given by
\begin{equation}
P_{\mu }=\int \int \int \mathcal{J}_{\mu }^{0}dx^{1}dx^{2}dx^{3}.
\end{equation}
With the use of Gauss' theorem one gets
\begin{equation}\label{M_P}
P_{\mu }=\frac{1}{8\pi }\int\int M_{\mu }^{0i}n_{i}dS.
\end{equation}

\section{Energy and Momentum Distributions for the Charged Regular Black Hole}
In order to perform the calculations using the Einstein energy-momentum
complex, it is necessary to transform the metric given by the line element
(\ref{line_element}) in Schwarzschild Cartesian coordinates. Thus, we obtain the line element
in the form:
\begin{equation}\label{line_element_SC}
ds^{2} =B(r)dt^{2}-(dx^{2}+dy^{2}+dz^{2})-\frac{B^{-1}(r)-1}{r^{2}}(xdx+ydy+zdz)^{2}\text{.}
\end{equation}
Now, after computing the superpotentials in the Einstein prescription we get
the following non-vanishing components:
\begin{equation}\label{h001}
h_{0}^{01}=\frac{2x}{r^{2}}\frac{2M}{r}\left\{1-\frac{1}{[1+(\frac{2Mr}{q^{2}})^{3}]^{(\alpha -3)/3}}\right\},
\end{equation}
\begin{equation}\label{h002}
h_{0}^{02}=\frac{2y}{r^{2}}\frac{2M}{r}\left\{1-\frac{1}{[1+(\frac{2Mr}{q^{2}})^{3}]^{(\alpha -3)/3}}\right\},
\end{equation}
\begin{equation}\label{h003}
h_{0}^{03}=\frac{2z}{r^{2}}\frac{2M}{r}\left\{1-\frac{1}{[1+(\frac{2Mr}{q^{2}})^{3}]^{(\alpha -3)/3}}\right\}.
\end{equation}
Using the line element (\ref{line_element}), with the metric coefficient (\ref{metric_coef_B}), the equation (\ref{Einstein_P}) for the energy and the expressions (\ref{h001})-(\ref{h003}) for the superpotentials, we obtain the energy distribution for the  regular charged black hole in the Einstein prescription:
\begin{equation}\label{Einstein_Energy}
E_{E}=M\left\{1-\frac{1}{[1+(\frac{2Mr}{q^{2}})^{3}]^{(\alpha -3)/3}}\right\}.
\end{equation}
Further, all the momenta are zero.

In Figures 2, 3 we can see the graph of the energy distribution as a function of the distance for different values of $\alpha$, while Figure 4 presents the energy distribution as a function of charge and mass. 

\begin{figure}[h!]
\includegraphics[width=84mm]{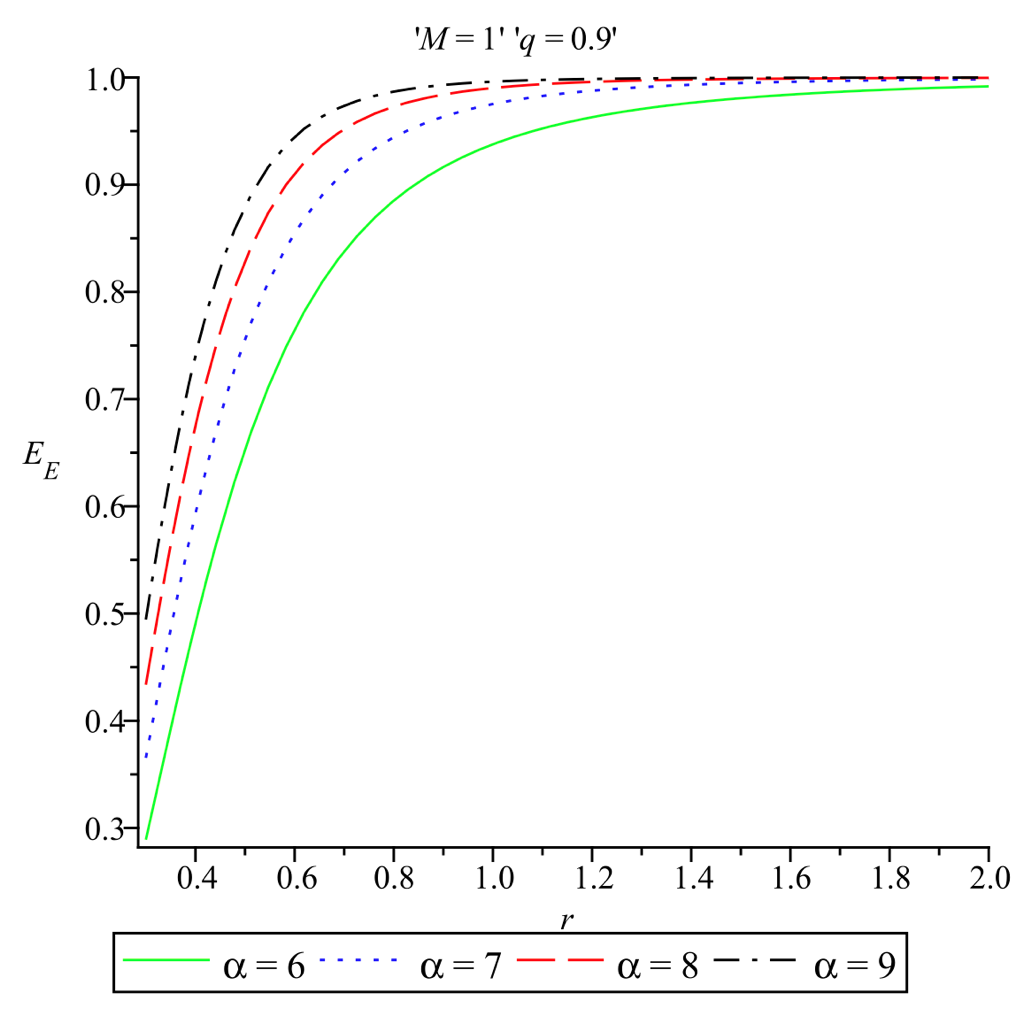}
\caption{Energy distribution calculated by the Einstein prescription between the inner and outer horizons.}
\label{fig5}
\end{figure}

\begin{figure}[h!]
\includegraphics[width=84mm]{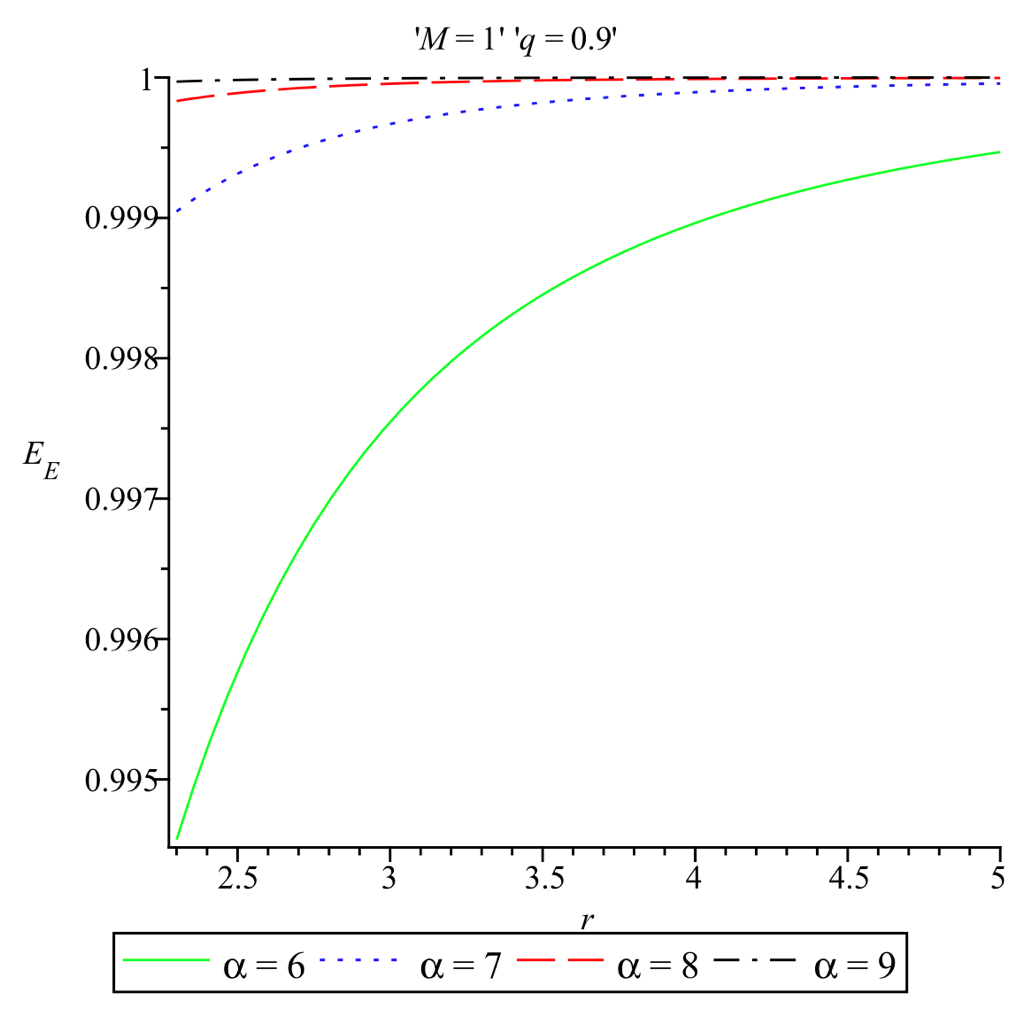}
\caption{Energy distribution calculated by the Einstein prescription outside the outer horizon.}
\label{fig6}
\end{figure}

\begin{figure}[h!]
\includegraphics[width=84mm]{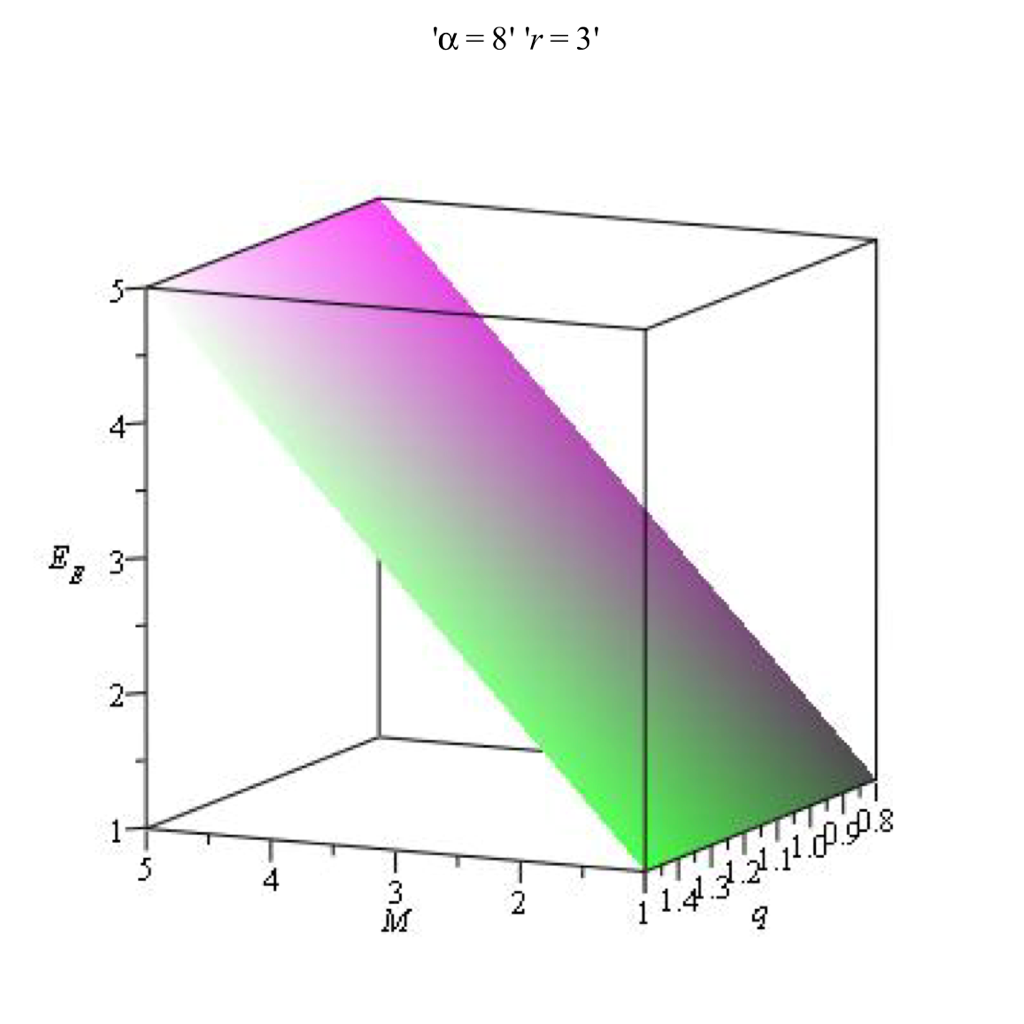}
\caption{Variation of the energy distribution calculated by the Einstein prescription with respect to charge and mass outside the outer horizon.}
\label{fig1}
\end{figure}

In order to apply the Landau-Lifshitz prescription, we use the $U^{\mu 0i}$ quantities defined in Eq.(\ref{LL_P}) to compute the energy--momentum distribution. 
The nonvanishing components of the $U^{\mu 0i}$ quantities are:
\begin{equation}\label{Uttx}
U^{ttx}=\frac{2x}{r^{2}}
\frac{\frac{2M}{r}\left[1-\frac{1}{[1+(\frac{2Mr}{q^{2}})^{3}]^{(\alpha -3)/3}}\right]}
{1-\frac{2M}{r}\left[1-\frac{1}{[1+(\frac{2Mr}{q^{2}})^{3}]^{(\alpha -3)/3}}\right]},
\end{equation}
\begin{equation}\label{Utty}
U^{tty}=\frac{2y}{r^{2}}
\frac{\frac{2M}{r}\left[1-\frac{1}{[1+(\frac{2Mr}{q^{2}})^{3}]^{(\alpha -3)/3}}\right]}
{1-\frac{2M}{r}\left[1-\frac{1}{[1+(\frac{2Mr}{q^{2}})^{3}]^{(\alpha -3)/3}}\right]},
\end{equation}
\begin{equation}\label{Uttz}
U^{ttz}=\frac{2z}{r^{2}}
\frac{\frac{2M}{r}\left[1-\frac{1}{(1+(\frac{2Mr}{q^{2}})^{3})^{(\alpha -3)/3}}\right]}
{1-\frac{2M}{r}\left[1-\frac{1}{[1+(\frac{2Mr}{q^{2}})^{3}]^{(\alpha -3)/3}}\right]}.
\end{equation}
Substituting (\ref{Uttx})-(\ref{Uttz}) in (\ref{LL_P}), we get the energy distribution:
\begin{equation}\label{LL_Energy}
E_{LL}=\frac{M\left[1-\frac{1}{[1+(\frac{2Mr}{q^{2}})^{3}]^{(\alpha -3)/3}}\right]}
{1-\frac{2M}{r}\left[1-\frac{1}{[1+(\frac{2Mr}{q^{2}})^{3}]^{(\alpha-3)/3}}\right]}.
\end{equation}
Also, for this prescription  all the momenta are found to vanish.

In the case of the Weinberg prescription we calculate the non-vanishing
components:
\begin{equation}\label{Dxtt}
D^{xtt}=\frac{2x}{r^{2}}
\frac{\frac{2M}{r}\left[1-\frac{1}{[1+(\frac{2Mr}{ q^{2}})^{3}]^{(\alpha -3)/3}}\right]}
{1-\frac{2M}{r}\left[1-\frac{1}{[1+(\frac{2Mr}{q^{2}})^{3}]^{(\alpha -3)/3}}\right]},
\end{equation}
\begin{equation}\label{Dytt}
D^{ytt}=\frac{2y}{r^{2}}
\frac{\frac{2M}{r}\left[1-\frac{1}{[1+(\frac{2Mr}{q^{2}})^{3}]^{(\alpha -3)/3}}\right]}
{1-\frac{2M}{r}\left[1-\frac{1}{[1+(\frac{2Mr}{q^{2}})^{3}]^{(\alpha -3)/3}}\right]},
\end{equation}
\begin{equation}\label{Dztt}
D^{ztt}=\frac{2z}{r^{2}}
\frac{\frac{2M}{r}\left[1-\frac{1}{[1+(\frac{2Mr}{q^{2}})^{3}]^{(\alpha -3)/3}}\right]}
{1-\frac{2M}{r}\left[1-\frac{1}{[1+(\frac{2Mr}{q^{2}})^{3})^{(\alpha -3)/3}}\right]}.
\end{equation}
Inserting the expressions obrained in (\ref{Dxtt})-(\ref{Dztt}) into (\ref{W_P}) we obtain for the
energy distribution inside a $2$-sphere of radius $r$:
\begin{equation}\label{W_Energy}
E_{W}=M\frac{\left[1-\frac{1}{[1+(\frac{2Mr}{q^{2}})^{3}]^{(\alpha -3)/3}}\right]}
{1-\frac{2M}{r}\left[1-\frac{1}{[1+(\frac{2Mr}{q^{2}})^{3}]^{(\alpha -3)/3}}\right]},
\end{equation}
while all the momenta vanish. As it can be seen, the energy in the Weinberg prescription (\ref{W_Energy}) is identical with the energy in the Landau-Lifshitz prescription (\ref{LL_Energy}).

In Figures 5, 6 we can see the graph of the energy distribution as a function of the distance for different values of $\alpha$, while Figure 7 presents the energy distribution as a function of charge and mass. 

\begin{figure}[h!]
\includegraphics[width=84mm]{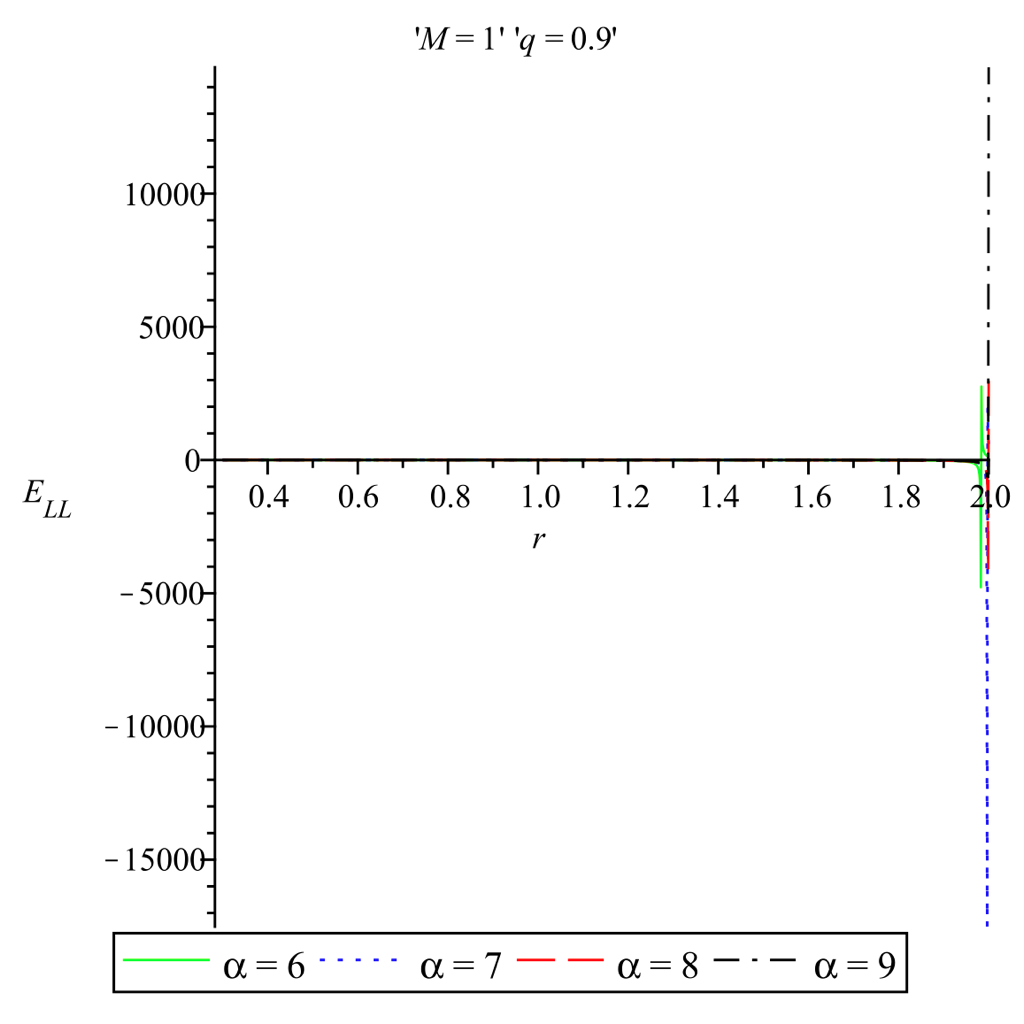}
\caption{Energy distribution calculated by the Landau-Lifshitz prescription between the inner and outer horizons.}
\label{fig8}
\end{figure}

\begin{figure}[h!]
\includegraphics[width=84mm]{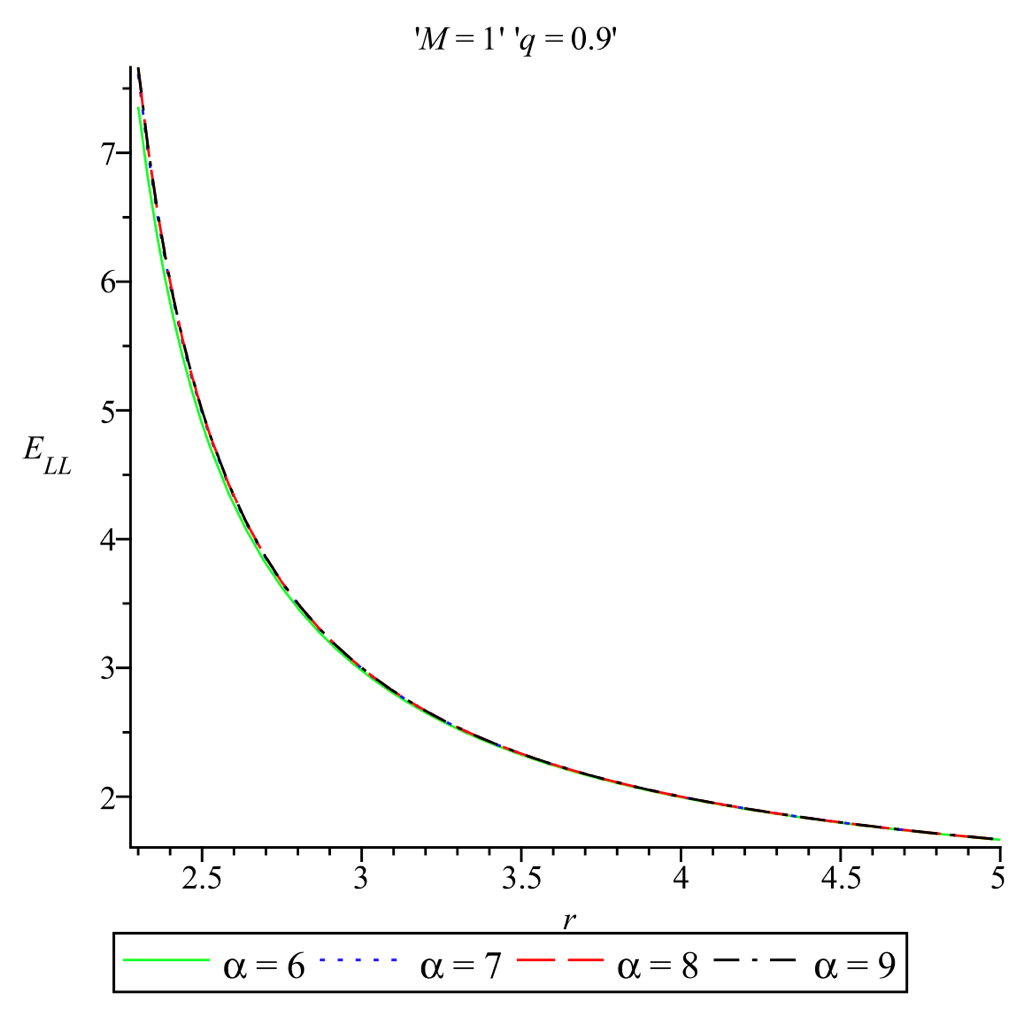}
\caption{Energy distribution calculated by the Landau-Lifshitz prescription outside the outer horizon.}
\label{fig9}
\end{figure}

\begin{figure}[h!]
\includegraphics[width=84mm]{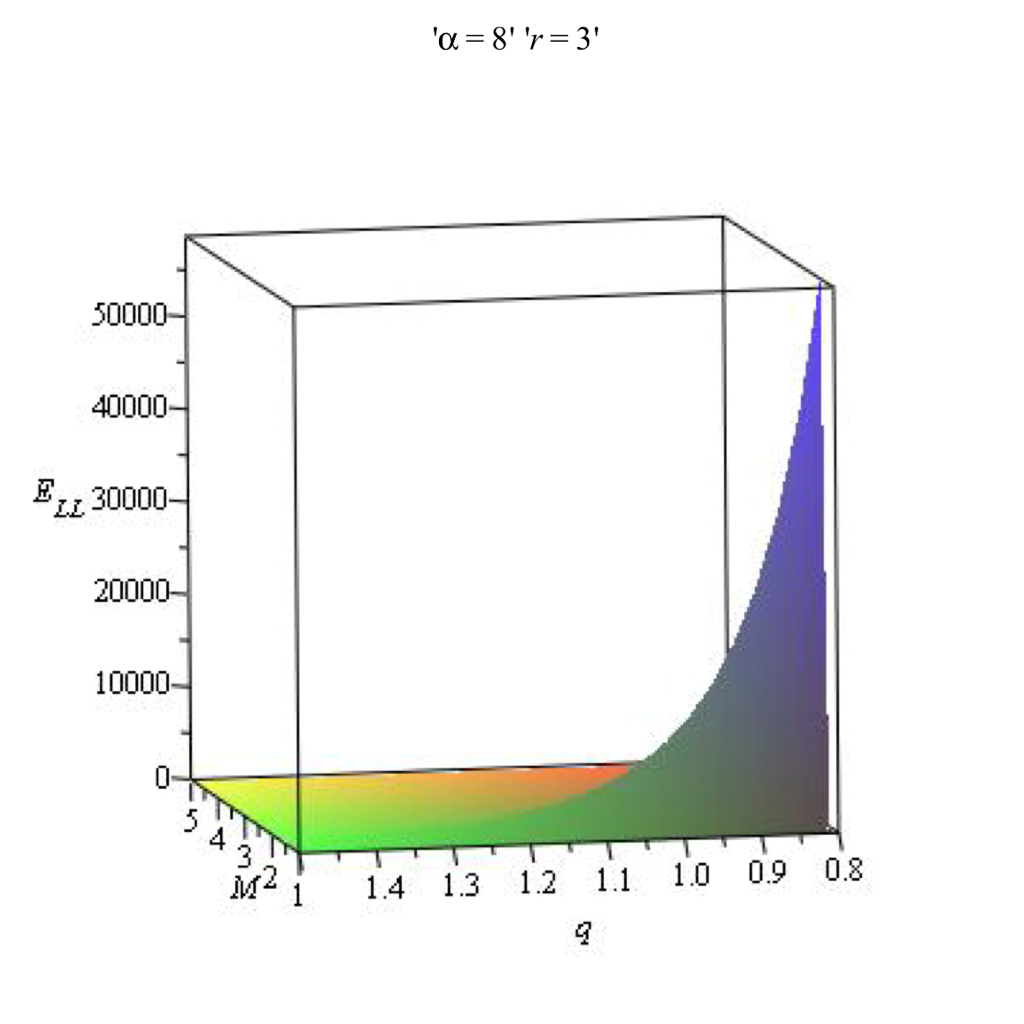}
\caption{Variation of the energy distribution calculated by the Landau-Lifshitz prescription with respect to charge and mass outside the outer horizon.}
\label{fig2}
\end{figure}

In the M\o ller prescription, the only non-vanishing superpotential is:
\begin{equation}\label{M001}
M_{0}^{01}=\left\{\frac{2M\left[1-\frac{1}{(1+\frac{8M^{3}r^{3}}{q^{6}})^{(\frac{\alpha }{3}-1)}}\right]}
{r^{2}}-\frac{48M^{4}(\frac{\alpha }{3}-1)r}{\left[1+\frac{8M^{3}r^{3}}{q^{6}}\right]^{(\frac{\alpha }{3}-1)}q^{6}\left[1+\frac{8M^{3}r^{3}}{q^{6}}\right]}\right\}r^{2}\sin \theta.
\end{equation}

Applying the last result for the line element (\ref{line_element}) with the metric coefficient (\ref{metric_coef_B}) and using the expression (\ref{M_P}), we get the energy distribution in the M\o ller prescription:
\begin{equation}\label{Moller_Energy}
E_{M}=\frac{r^{2}}{2}
\left\{\frac{2M\left[1-\frac{1}{(1+\frac{8M^{3}r^{3}}{q^{6}})^{(\frac{\alpha }{3}-1)}}\right]}{r^{2}}-
\frac{48M^{4}(\frac{\alpha }{3}-1)r}{\left[1+\frac{8M^{3}r^{3}}{q^{6}}\right]^{(\frac{\alpha }{3}-1)}q^{6}\left[1+\frac{8M^{3}r^{3}}{q^{6}}\right]}\right\}.
\end{equation}
Also, in this prescription it is found that  all the momenta vanish.

In Figures 8, 9 we can see the graph of the energy distribution as a function of the distance for different values of $\alpha$, while Figure 10 presents the energy distribution as a function of charge and mass. 

\begin{figure}[t!]
\includegraphics[width=84mm]{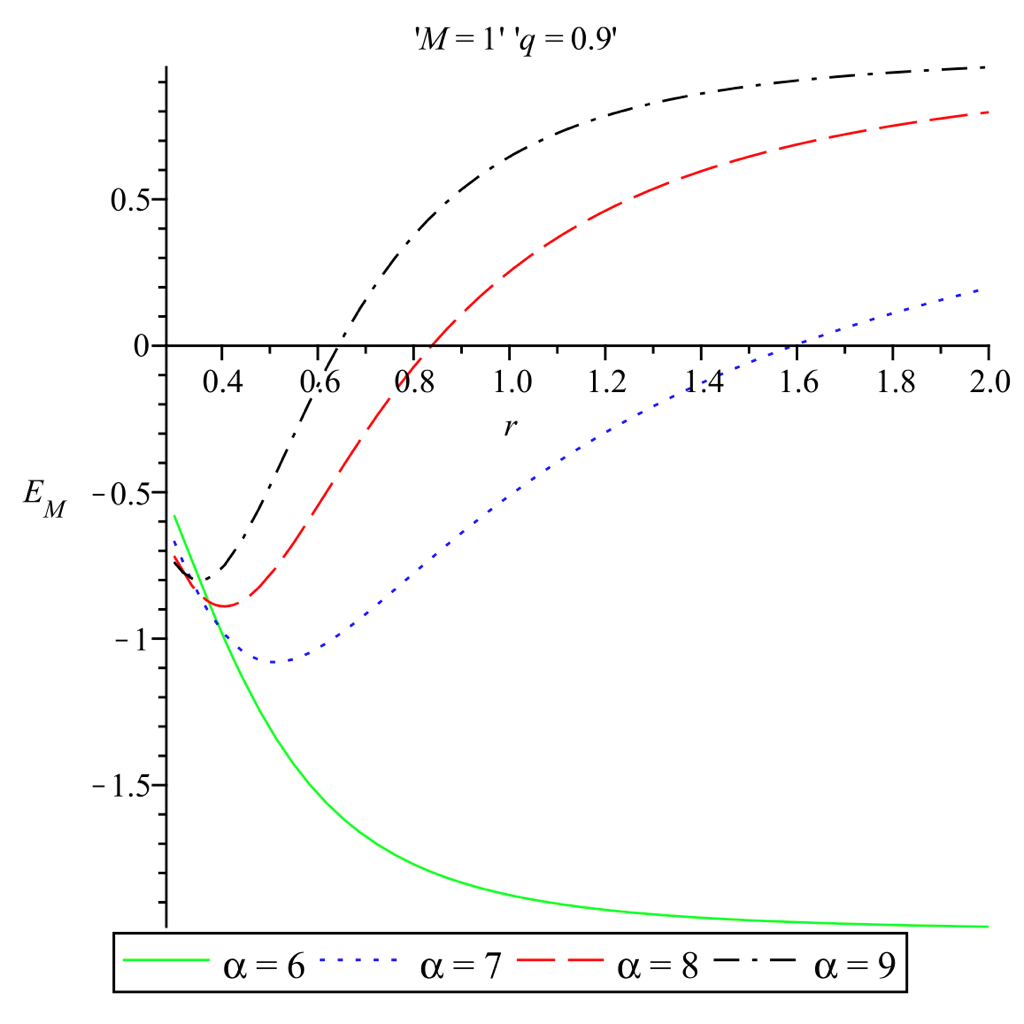}
\caption{Energy distribution calculated by the M\o ller prescription between the inner and outer horizons.}
\label{fig10}
\end{figure}

\begin{figure}[t!]
\includegraphics[width=84mm]{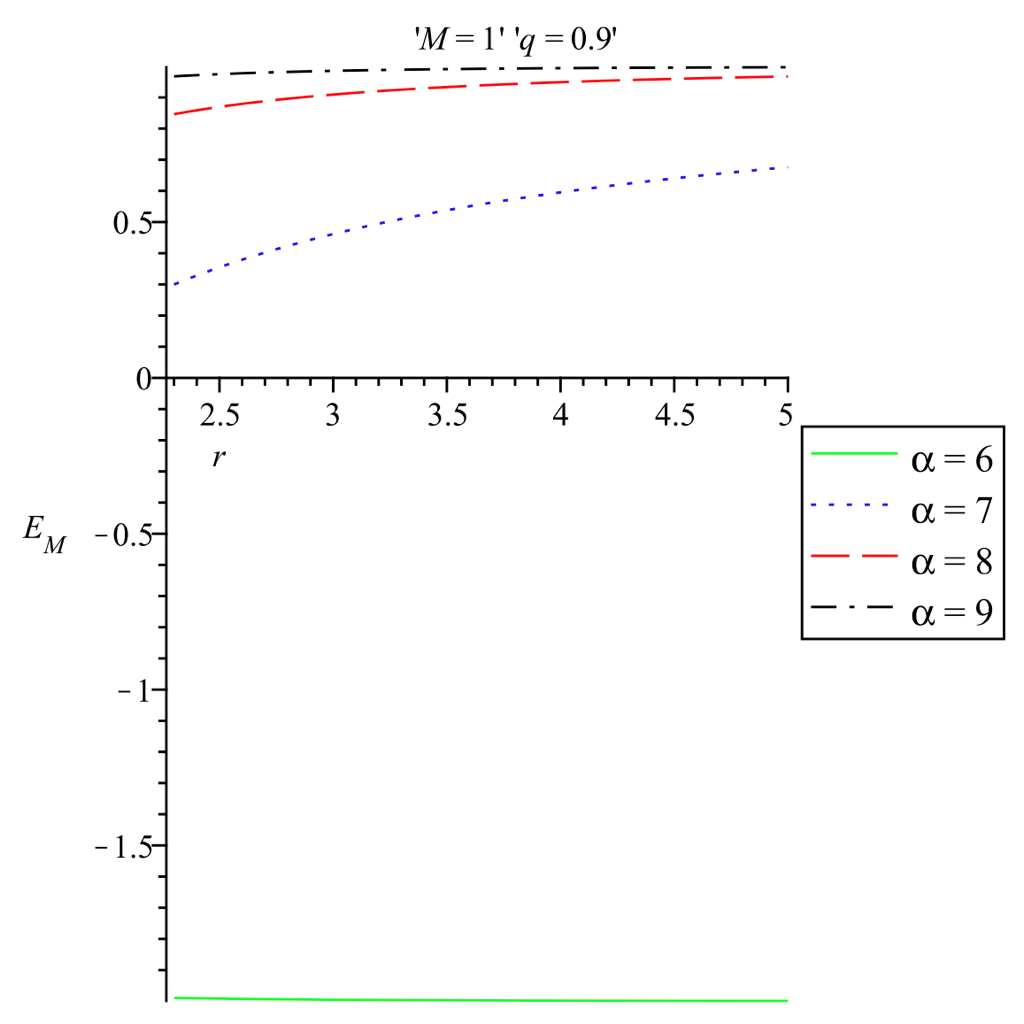}
\caption{Energy distribution calculated by the M\o ller prescription outside the outer horizon.}
\label{fig11}
\end{figure}

\begin{figure}[t!]
\includegraphics[width=84mm]{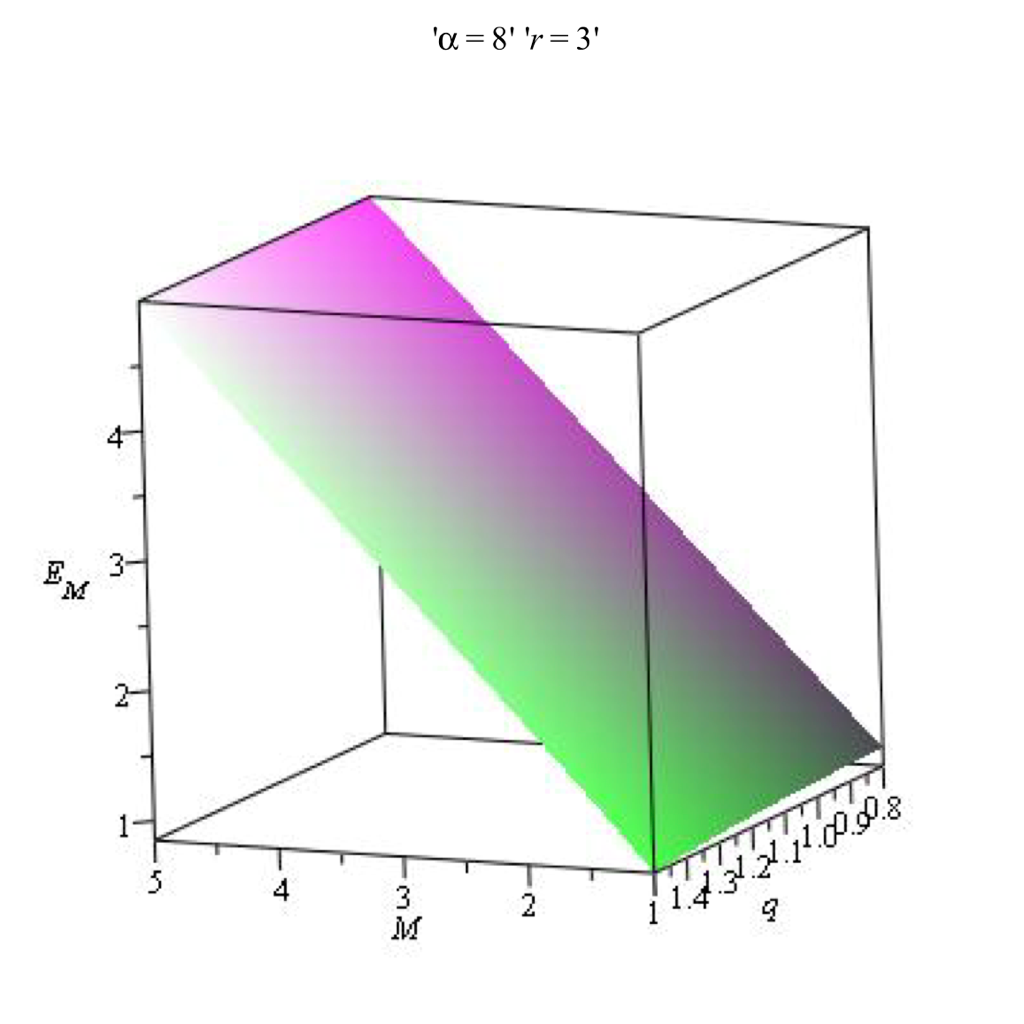}
\caption{Variation of the energy distribution calculated by the M\o ller prescription with respect to charge and mass outside the outer horizon.}
\label{fig3}
\end{figure}

It is pointed out, according to Figure 8, that for $\alpha \leqslant  6$ the M\o ller energy of the regular black hole is negative and decreases monotonically for the whole range of the values of $r$ considered.

Finally, in Figure 11 we give a comparison of the energy distributions as obtained by the different prescriptions for $\alpha=8$.
\begin{figure}[h!]
\includegraphics[width=84mm]{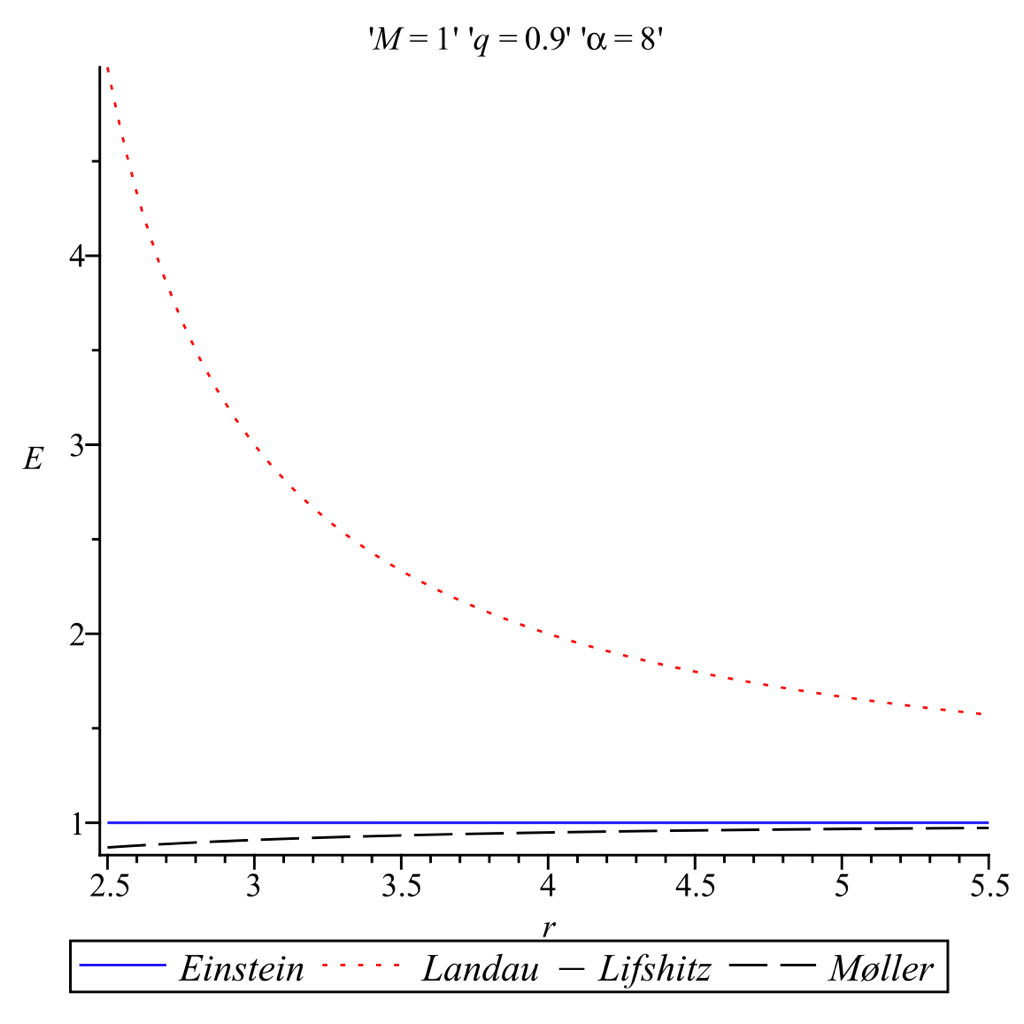}
\caption{Comparison of the energy distributions calculated by the Einstein, Landau-Lifshitz and M\o ller prescriptions outside the outer horizon.}
\label{fig4}
\end{figure}

\section{Discussion}

This paper is focused on the study of the energy-momentum for the gravitational field of a new four-dimensional, spherically symmetric, static and charged regular black hole solution developed
in the context of Einstein-nonlinear electrodynamics coupling, by applying the Einstein,
Landau-Lifshitz, Weinberg and M\o ller energy-momentum complexes. The metric considered, which has two horizons, does not asymptotically behave as the Reissner-Nordstr\"{o}m black hole
solution unless $\mu =4$, where $\mu $ is a positive integer parameter.
The expressions for the energy are well-defined in all the aforesaid prescriptions and
depend on the mass $M$ of the black hole, its charge $q$, a positive integer $\alpha $ and the radial coordinate $r$. Both the Landau-Lifshitz and Weinberg
prescriptions give the same result for the energy distribution, while all the momenta are zero in all four prescriptions used.

In order to examine the physical meaning of the results obtained, we study the limiting
behavior of the energy for $r\rightarrow \infty$, $r=0$ and $q=0$ in three
cases: (i) $1\leqslant \alpha <3$, (ii) $\alpha =3$ and (iii) $\alpha >3$. However, since the cases (i) and (ii) do not yield any physically meaningful result, we present only the results for the case (iii) in Table 1.

\begin{table}[t!]
\centering
\begin{tabular}{|l|c|c|c|}
\hline
Prescription & $r\rightarrow \infty$ & $r=0$ & $q=0$ \\ 
\hline
&&&\\
Einstein & $M$ & $0$ & $M$ \\ 
\hline
&&&\\
Landau-Lifshitz & $M$ & $0$ & $M(1-\frac{2M}{r})^{-1}$ \\
\hline
&&&\\
Weinberg & $M$ & $0$ & $M(1-\frac{2M}{r})^{-1}$ \\ 
\hline
&&&\\
M\o ller & $M$ & $0$ & $M$\\
\hline
\end{tabular}
\caption{Limiting cases for $\alpha>3$}
\end{table}

For $\mu =4$ and $\alpha=3$ we calculate the energy-momentum distribution for the asymprotically Reissner-Nordstr\"{o}m metric with the metric coefficient (\ref{RN_solution}).

For this black hole solution, the energy distribution obtained by using the Einstein, Landau-Lifshitz, Weinberg and M\o ller energy-momentum complexes is given in the following Table 2, along with the limiting cases for $r\rightarrow \infty $, $r=0$ and $q=0$.

\begin{table}[t!]
\centering
\begin{tabular}{|l|c|c|c|c|}
\hline
\textbf{Prescription} & \textbf{Energy} &$r\rightarrow \infty $ & $r=0$ & $q=0$ \\ 
\hline
&&&&\\
Einstein & $M\left\{1-\frac{1}{[1+(\frac{2Mr}{q^{2}})^{3}]^{1/3}}\right\}$ & $M$ & $0$ & $M$ \\
&&&&\\
\hline
&&&&\\ 
Landau-Lifshitz & $\frac{M\left[(1-\frac{1}{[1+(\frac{2Mr}{q^{2}})^{3}]^{1/3}}\right]}
{1-\frac{2M}{r}\left[1-\frac{1}{[1+(\frac{2Mr}{q^{2}})^{3}]^{1/3}}\right]}$ & $M$ & $0$ & $M((1-\frac{2\,M}{r})^{-1}$\\ 
&&&&\\
\hline
&&&&\\
Weinberg & $\frac{M\left[1-\frac{1}{[1+(\frac{2Mr}{q^{2}})^{3}]^{1/3}}\right]}
{1-\frac{2M}{r}\left[1-\frac{1}{[1+(\frac{2Mr}{q^{2}})^{3}]^{1/3}}\right]}$ & $M$ & $0$ & $M((1-\frac{2\,M}{r})^{-1}$\\
&&&&\\
\hline
&&&&\\ 
M\o ller & $\frac{r^{2}}{2}
\left\{-\frac{2M\left[(\frac{8\,M^{3}r^{3}}{q^{6}}+1)^{-\frac{1}{3}}-1\right]}{r^{2}}-
\frac{16M^{4}r}{q^{6}\left[\frac{8M^{3}r^{3}}{q^{6}}+1\right]^{\frac{4}{3}}}\right\}$& $M$ & $0$ & $M$\\
&&&&\\
\hline
\end{tabular}
\caption{$\mu=4$ and $\alpha=3$}
\end{table}

As we can see, the limiting behavior of the energy for the asymptotically Reissner-Nordstr\"{o}m solution 
($\mu=4$, $\alpha=3$) is the same as in the case of the non-asymptotically Reissner-Nordstr\"{o}m solution given in Table 1.

Thus, by comparing the results obtained for the energy with all the energy-momentum complexes applied, we see that, for $q=0$ and at infinity, the Einstein prescription and the M\o ller prescription give the same results (namely, the ADM mass $M$) with that obtained for the Schwarzschild black hole solution. On the other hand, the Landau-Lifshitz and the Weinberg prescriptions, while at infinity lead to the same result (the ADM mass $M$), in the chargeless case $q=0$ give the energy distribution  $E_{LL}=E_{W}=M(1-\frac{2M}{r})^{-1}$ which is in total agreement with the expression obtained for the Schwarzschild black hole using Schwarzschild Cartesian coordinates (see, K.S. Virbhadra (1999), in \cite{Aguirregabiria}).

Finally,  all the four prescriptions used yield a vanishing energy at $r=0$. 

Based on the present study, as well as on the aforementioned agreement with previous results
obtained in the case of the Schwarzschild metric, we conclude that the Einstein and M\o ller
energy-momentum complexes give physically meaningful results and, thus,
can be considered as the most reliable tools among the energy-momentum complexes for the study of the energy-momentum localization of gravitational systems.

An interesting future perspective lies in the calculation of the 
energy-momentum of the new regular black hole solution considered in the present work by applying  other energy-momentum complexes and/or the tele-parallel equivalent to general relativity (TEGR).


\begin{thebibliography}{99}
\bibitem{Bel} L. Bel, C. R. Acad. Sci. Paris \textbf{246}, 3105 (1958);
 I. Robinson, Class. Quantum Grav. \textbf{14}, A331 (1997); 
 M.A.G. Bonilla and J.M.M. Senovilla, Gen. Rel. Grav. \textbf{29}, 91 (1997); 
 J.M.M. Senovilla, Class. Quantum Grav. \textbf{17}, 2799 (2000).

\bibitem{Brown} J. D. Brown and J.W. York, Phys. Rev. D\textbf{47}, 1407 (1993);
Sean A. Hayward, Phys. Rev. D\textbf{49}, 831 (1994); 
C. M. Chen, J. M.  Nester, Class. Quantum Grav. \textbf{16}, 1279 (1999); 
C-C.M. Liu and S. T. Yau, Phys. Rev. Lett. \textbf{90}, 231102 (2003); 
L. Balart, Phys. Lett. B\textbf{687}, 280 (2010)

\bibitem{Einstein} A. Einstein, Preuss. Akad. Wiss. Berlin \textbf{47}, 778 (1915);
Addendum-ibid. \textbf{47}, 799 (1915); 
A. Trautman, in \textit{Gravitation: an Introduction to Current Research}, L. Witten (ed.), Wiley, New York, p. 169 (1962).

\bibitem{Landau} L. D. Landau and E.M. Lifshitz, \textit{The Classical Theory of Fields}, Pergamon Press, p. 280 (1987).

\bibitem{Papapetrou} A. Papapetrou, Proc. R. Irish. Acad. A\textbf{52}, 11 (1948).

\bibitem{Bergmann} P. G. Bergmann and R. Thomson, Phys. Rev. \textbf{89}, 400 (1953).

\bibitem{Moller} C. M\o ller, Ann. Phys. (NY) \textbf{4}, 347 (1958).

\bibitem{Weinberg} S. Weinberg, \textit{Gravitation and Cosmology: Principles and
Applications of General Theory of Relativity}, John Wiley and Sons Inc., New York,  p. 165 (1972).

\bibitem{Qadir} A. Qadir and M. Sharif, Phys. Lett. A\textbf{167}, 331 (1992).

\bibitem{Moller_1} C. M\o ller, Nucl. Phys. \textbf{57}, 330 (1964); 
K. Hayashi and T. Shirafuji, Phys. Rev. D\textbf{19}, 3524 (1979); 
J.M. Nester, Lau Loi So and T. Vargas, Phys. Rev. D\textbf{78}, 044035 (2008); 
Gamal G. L. Nashed and T. Shirafuji, Int. J. Mod. Phys. D\textbf{16}, 65 (2007); 
Gamal G. L. Nashed, Chin. Phys. Lett. \textbf{25}, 1202 (2008); 
J.W. Maluf, F.F. Faria and S.C. Ulhoa, Class. Quantum Grav. \textbf{24}, 2743 (2007); 
J.W. Maluf, M.V.O. Veiga and J.F. da Rocha-Neto, Gen. Rel. Grav. \textbf{39}, 227 (2007); 
A.A. Sousa, R.B. Pereira, A.C. Silva, Grav. Cosmol. \textbf{16}, 25 (2010); 
M. Sharif, Sumaira Taj, Astrophys. Space Sci. \textbf{325}, 75 (2010);
Liu Y.X., Zhao Z.H., Yang J., and Duan Y.S., arXiv:0706.3245 [gr-qc].

\bibitem{Virbhadra} K. S. Virbhadra, Phys. Rev. D\textbf{41}, 1086 (1990); 
K. S. Virbhadra, Phys. Rev. D\textbf{42}, 2919 (1990); 
N. Rosen and K.S. Virbhadra, Gen. Rel. Grav. \textbf{25}, 429 (1993); 
K. S.Virbhadra and J. C. Parikh, Phys. Lett. B\textbf{331}, 302 (1994);  
S. S. Xulu, Mod. Phys. Lett. A\textbf{15},1511 (2000); 
I-Ching Yang and I. Radinschi, Chin. J. Phys., \textbf{42(1)}, 40 (2004); 
I. Radinschi, Th. Grammenos, Int. J. Theor. Phys., \textbf{47(5)}, 1363 (2008); 
I-Ching Yang, Chi-Long Lin, Irina Radinschi, Int. J. Theor. Phys., \textbf{48(1)}, 248 (2009); 
I. Radinschi, F. Rahaman and A. Ghosh, Int. J. Theor. Phys. \textbf{49}, 943 (2010); 
I. Radinschi, F. Rahaman, A. Banerjee, Int. J. Theor. Phys., \textbf{50(9)}, 2906 (2011); 
Ragab M. Gad, Int. J. Theor.Phys. \textbf{46}, 3263 (2007); 
M. Abdel-Megied, Ragab M. Gad, Adv. High Energy Phys. \textbf{2010}, 379473 (2010); 
Ragab M. Gad, Astrophys.Space Sci. \textbf{346,} 553 (2013); 
T. Bringley, Mod. Phys. Lett.  A\textbf{17}, 157 (2002); 
M. Sukenik and J. Sima, arXiv:grqc/0101026; 
M. Sharif and Tasnim Fatima, Astrophys. Space Sci. \textbf{302}, 217 (2006); 
M. Sharif, M. Azam, Int. J. Mod. Phys. A\textbf{22}, 1935 (2007); 
P. Halpern, Astrophys. Space. Sci. \textbf{306}, 279 (2006); 
E. C. Vagenas, Int. J. Mod. Phys. A\textbf{18}, 5781 (2003); 
E. C. Vagenas, Mod. Phys. Lett. A\textbf{19}, 213 (2004); 
E. C. Vagenas, Int. J. Mod. Phys. D\textbf{14}, 573 (2005); 
E. C. Vagenas, Mod. Phys. Lett. A\textbf{21}, 1947 (2006); 
T. Multamaki, A. Putaja, E. C. Vagenas and I. Vilja, Class. Quant. Grav. \textbf{25}, 075017 (2008); 
L. Balart, Mod. Phys. Lett. A\textbf{24}, 2777 (2009); 
Amir M. Abbassi, Saeed Mirshekari, Amir H. Abbassi, Phys. Rev. D\textbf{78}, 064053 (2008); 
J. Matyjasek, Mod. Phys. Lett. A\textbf{23(8)}, 591 (2008).

\bibitem{Aguirregabiria} J. M. Aguirregabiria, A. Chamorro and K. S. Virbhadra, Gen. Rel. Grav. 
\textbf{28}, 1393 (1996); 
K. S. Virbhadra, Phys. Rev. D\textbf{60}, 104041 (1999);
S. S. Xulu, Int. J. Theor. Phys. \textbf{46}, 2915 (2007).

\bibitem{Penrose} R. Penrose, Proc. R. Soc. London A\textbf{381}, 53 (1982).

\bibitem{Tod} K.P. Tod, Proc. R. Soc. London A\textbf{388}, 457 (1983).

\bibitem{Gamal} Gamal G. L. Nashed, Mod. Phys. Lett. A\textbf{22}, 1047 (2007);
Gamal G. L. Nashed, Chin. Phys. Lett. \textbf{25}, 1202 (2008); Gamal G. L.
Nashed, Int. J. Mod. Phys. A\textbf{23}, 1903 (2008); Gamal G.L. Nashed,
Chin. Phys. B\textbf{19(2)}, 020401 (2010); 
M. Sharif, Abdul Jawad, Astrophys. Space Sci. \textbf{331}, 257 (2011).

\bibitem{Chang} Chia-Chen Chang, J. M. Nester and Chiang-Mei Chen, Phys. Rev.
Lett.\textbf{\ 83}, 1897 (1999); 
J. M. Nester, Chiang-Mei Chen, Jian-Liang
Liu, Gang Sun, in \textit{Relativity and Gravitation - 100 years after Einstein in Prague}, J. Bicak and T. Ledvinka (eds.), Springer Proceedings in Physics, Vol. 157, p.177 (2014).

\bibitem{Bardeen} J.M. Bardeen, in Proc. of the Intern. Conf. GR5, Tbilisi, USSR, p.174 (1968).

\bibitem{Ansoldi} S. Ansoldi, arXiv:0802.0330

\bibitem{Ayon} E. Ay\'on-Beato and A. Garc\'{\i}a,  Phys. Rev. Lett. \textbf{80}, 5056 (1998)

\bibitem{Beato} E. Ay\'on-Beato and A. Garc\'{\i}a, Phys. Lett. B\textbf{464},  25 (1999);
E. Ay\'on-Beato and A. Garc\'{\i}a,  Gen. Rel. Grav. \textbf{31},  629 (1999);
M. Cataldo and A. Garc\'{\i}a,  Phys. Rev. D\textbf{61},  084003 (2000); 
K.A. Bronnikov, Phys. Rev. D\textbf{63}, 044005 (2001);
A. Burinskii and S.R. Hildebrandt, Phys. Rev. D\textbf{65}, 104017 (2002);
J. Matyjasek, Phys. Rev. D\textbf{70}, 047504 (2004);
E. Ay\'on-Beato and A. Garc\'{\i}a,  Gen. Rel. Grav. \textbf{37}, 635 (2005);  
I-Ching Yang, Chi-Long Lin and I. Radinschi,   Int. J. Theor. Phys. \textbf{48}, 248 (2009);
J. Matyjasek, D. Tryniecki and M. Klimek, Mod. Phys. Let. A\textbf{23}, 3377 (2009);
S.H. Hendi,  Ann. Phys. (N.Y.) \textbf{333}, 282 (2013)

\bibitem{Garcia} E. Ay\'on-Beato and A. Garc\'{\i}a,  Phys.Lett. B\textbf{493},  149 (2000)

\bibitem{Balart} L. Balart and E.C. Vagenas, Phys. Lett. \textbf{B14}, 730 (2014).

\bibitem{Hayward} S.A. Hayward, Phys. Rev. Lett. \textbf{96}, 031103 (2006)

\end{thebibliography}
\end{document}